\documentclass[preprint,12pt]{elsarticle}

\usepackage{graphicx}
\usepackage{color}

\usepackage[cmex10]{amsmath}
\usepackage{wrapfig}

\journal{a peer-reviewed journal}

\begin{document}

\begin{frontmatter}



\title{A Prototype of Wireless Power and Data Acquisition System for Large Detectors}


\author[anl]{P.~De Lurgio}
\author[anl]{Z.~Djurcic}
\author[anl]{G.~Drake}
\author[niu]{R.~Hashemian}
\author[anl]{A.~Kreps}
\author[anl]{M. Oberling}
\author[niu]{T.~Pearson} 
\author[anl]{H.~Sahoo}

\address[anl]{Argonne National Laboratory, Argonne, IL 60439, USA}
\address[niu]{Northern Illinois University, Dekalb, IL 60115, USA}

\begin{abstract}

A new prototype wireless data acquisition system has been developed with the intended application to read-out 
instrumentation systems having a large number of channels. In addition such system could be deployed in smaller 
detectors requiring increased mobility.
The data acquisition and control system is based on 802.11n compliant hardware and protocols.
In this paper we describe our case study with a single readout channel
performed for a potential large detector containing photomultiplier tubes. The front-end circuitry, including a high-voltage power 
supply is powered wirelessly thus creating an all-wireless detector readout. The benchmarked performance of the prototype 
system and how a large scale implementation of the system might be realized are discussed.

\end{abstract}


\end{frontmatter}


\section{Introduction}
\label{sec_intro}

The goal of this R\&D project was to build a detector module that operates from wireless power and sends data wirelessly. 
The motivation is the elimination of the massive cable plants that are typical of large detectors, which can result in cost reductions, 
simplified installation and repair, and reductions in detector  dead mass. 
Specially, cabling is not practical for detectors in remote locations or hostile environments.
The primary purpose of the effort described here is to ascertain the feasibility and 
practicality of such devices as single detector modules configured as arrays in a large detector.
We chose a photomultiplier tube (PMT) as the basis for this study, in part because of ubiquity of PMTs in high energy physics detectors, as well as the compact nature of the detector. 
In particular we selected a 10 inch diameter PMT, a Hamamatsu R7081~\cite{ref_R7081}, which has a dark noise rate of $\sim$10 kHz.
The system specifications include the capability to measure single photoelectrons, which imposes requirements on bandwidth and sensitivity of the instrumentation. 
Such requirements are often listed in connection with large neutrino detectors
and include examples of considered water-based LBNE~\cite{wc_lbne}, MEMPHYS~\cite{memphys}, Hyper-K~\cite{HyperK}, and CHIPS~\cite{chips} detectors.
Detectors used for homeland security also use some of these technologies and techniques.	Simplifying the implementation 
of large detectors with high channel counts will facilitate optimized detection of nuclear materials and other contra-band. \\
Numerous additional applications could benefit from infrastructure simplification, increased mobility, and stand-off distance
capability of wireless techniques. Some examples of these include monitoring of the neutrino flux at a nuclear reactor, large mobile neutron detectors for detection of radioactive material in security applications, and detectors operating in high- radiation areas. Operation in real world conditions requires the ability for detectors to be easily relocated, and the ability to operate continuously for long periods of time, both of these are improved using low-power wireless techniques.\\
For neutrino and other low-rate experiments, the single photoelectron rate dominates the event rate. 
If we assume for each event that the data would be comprised of 6 bytes, 2 bytes of pulse height information and a 4 byte time-stamp, at 10 kHz, this translates to 60 kB, or 480 kb/s.
For a single detector element, these data rates are achievable for a wireless readout.
Extrapolating to a large detector with tens of thousands of channels translates into a data rate on the order of $\sim$20 Gb/s, which is quite challenging for wireless readout. 
The target specifications for the prototype system are summarized in Table~\ref{tab:table1}.\\
Successful demonstration of an all wireless system could be transformational when realized in a practical, cost effective, and reliable fashion.
This prototyping work could in principle, with more design effort, allow for applications of these new read-out technologies. 

\begin{table*}
\centering
\caption{\label{tab:table1} Target specifications for the wireless data acquisition system.}
\begin{tabular}{c|c}
     \hline\hline
Specification                                                    & Target        \cr   \hline 
Maximum event rate (single p.e.)                &   $10$ kHz \cr   \hline
Bytes per event                                               &  6 (2 pulse height, 4 time-stamp) \cr    \hline
Average data rate per front-end channel   &  $60$ kB/s   \cr \hline
Total power consumption @ 10 kHz                    & $250$ mW \cr
                        Digital                                       &   $120$ mW \cr
                     Front-end                                    &      $30$ mW \cr
                         HV                                            &      $80$ mW \cr \hline
                      Data transfer rate                     &      $35$ Mb/s \cr \hline
Bit error rate                                                    &      $<$ 1$\times$10$^{-12}$   \cr \hline
Additional Features                                       &      Self-trigger for pedestals \cr
                                                                          &     Data pull                            \cr
                                                                          &   Programmable HV                \cr
                                                                          &   Programmable Discriminator \cr    \hline\hline
\end{tabular}
\end{table*}

\section{Design Considerations}
\label{sec_design}

The R\&D project began by researching the different technologies for wireless data and power transmission.
For wireless data transmission, two techniques in use today, radio
frequency (RF) wireless and free-space optical were considered.
Similarly, for wireless power transmission RF and optical were considered as well.
Other methods of transmitting power wirelessly; for example,
through strongly coupled magnetic resonances~\cite{mit_paper} were considered.
In selecting technologies, a preference was placed on the use of inexpensive, off-the-shelf technologies
that could be implemented relatively easily, with minimal risk and which could meet the performance goals of this project.

\subsection{Wireless Data Transmission}

For wireless data transmission, optical links support higher data rates than RF.  Individual free space links over distances
of meters can achieve transfer rates of approximately 1 Gb/s~\cite{optical}. However, RF transmission does not require
line-of-sight, and an individual transmitter can communicate with many front-ends. RF data transmission was 
chosen for this project because both of these advantages provide significant simplification and cost reduction. 
The difficulty with RF is that the bandwidth of an individual link is typically modest, approximately 100 MB/s, 
and the available RF bandwidth is shared by all elements of the detector.

There are two primary categories of wireless data technologies in use today. These are mobile/cellular, and wireless 
local area network (WLAN). Each has different variants for specific applications. For this project, we focused on WLAN technologies, since they have the highest data throughput. Of the different WLAN variants, we selected 802.11n, since at the time,
it offered the highest data throughput and has sufficient range for this application. Combined with the ubiquity of hardware 
and the relative ease of implementation, made this technology the preferred option. 
The total data rate of a single steam 802.11n link is approximately 65 Mb/s~\cite{connect_blue}. 
Note that this is the total data rate and not the payload data rate, which is $\sim$35 Mb/s. 
This is sufficient for our single prototype front-end. However, for a large detector, the challenge would be in transferring data 
from thousands of readout channels over a limited and common frequency spectrum.

One frequency range in 802.11n is centered in the U-NII and ISM bands at about 5.5 GHz. It has an overall bandwidth 
of approximately 1.2 GHz from about 4.9 GHz to 6.1 GHz. This ignores limitations imposed by the governing agencies 
that dictate the use of the electromagnetic spectrum. Under certain circumstances, the full spectrum can be available for 
use with dynamic frequency selection capable devices~\cite{802_11}. The approach uses a large number of access points
in a large detector system to accommodate the total payload data rate needed by the data acquisition system. 
Single stream 802.11n access points can have an individual operating bandwidth of 20 MHz or 40 MHz. 
With 20 MHz wide channels, it should be possible to populate the 1.2 GHz frequency with up to 48 access points. 
For a single stream 802.11n link this translates into a total payload data rate of 1.68 Gb/s, as illustrated in Figure~\ref{fig:fig_wirelessdata} below.
The system consists of many access points ($\sim$48) each communicating to many clients ($\sim$64), for a total of 3072 clients. 
This assumes 480 kb/s per front-end and a burst transfer rate of 35 Mb/s.
%
%
%
\begin{figure}
	\resizebox{\columnwidth}{!}{\includegraphics{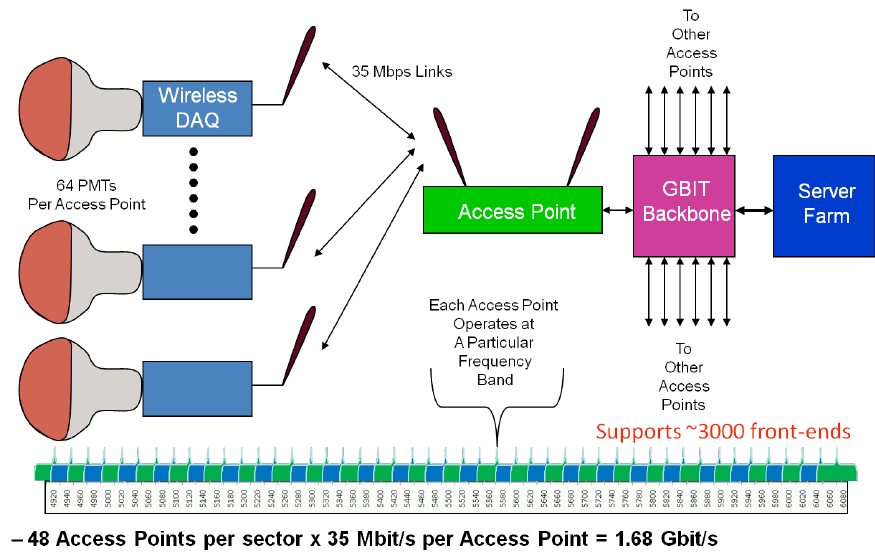}}
\caption{Allocation of frequency space for wireless data transmission.}
\label{fig:fig_wirelessdata} 
\end{figure}

For a large detector with tens of thousands of channels, this payload data rate is clearly insufficient. There are 
two extensions of this concept to increase the data throughput. First, the use of directional antennas and the intrinsic 
shielding created by the detector volume should allow the same frequency range to be utilized in different locations
on the detector. This effectively multiplies the usable spectrum, where the limits would depend on the particular
detector geometry. Second, the 802.11n standard includes extensions that allow for multiple input and output (MIMO)
wireless streams. This extension could significantly increase data throughput. For example, a $4\times4$ MIMO implementation 
supports 4 simultaneous transmit and receive streams to yield a data throughput of 600 Mb/s, which translates into 
a maximum payload data rate of 320 Mb/s. However, $4\times4$ MIMO requires 40 MHz wide channels, which means 
the 1.2 GHz frequency band would support approximately 24 access points. This translates into a much higher 
total ``sector'' payload data rate of 7.68 Gb/s. The combination of these two extensions to this multi-access point concept 
should provide sufficient data throughput to read-out a large detector.
Note that the above calculations do not account for latency in the communication. In practice, the latencies will impact
the achievable throughput. In a real system, the DAQ system would probably have to poll the front-ends for data.
This creates a defined handshake between client and access point and in principle increases latencies slightly but 
controls the readout process and avoids the need for clients requesting to send data. Ultimately this method increases 
the overall throughput. For our system, the intention is to have the server request data from each front-end 
with a periodicity of $1$s.

\subsection{Wireless Power Transmission}
\label{sec_powertransfer}

\begin{figure}[ht]
	\resizebox{\columnwidth}{!}{\includegraphics{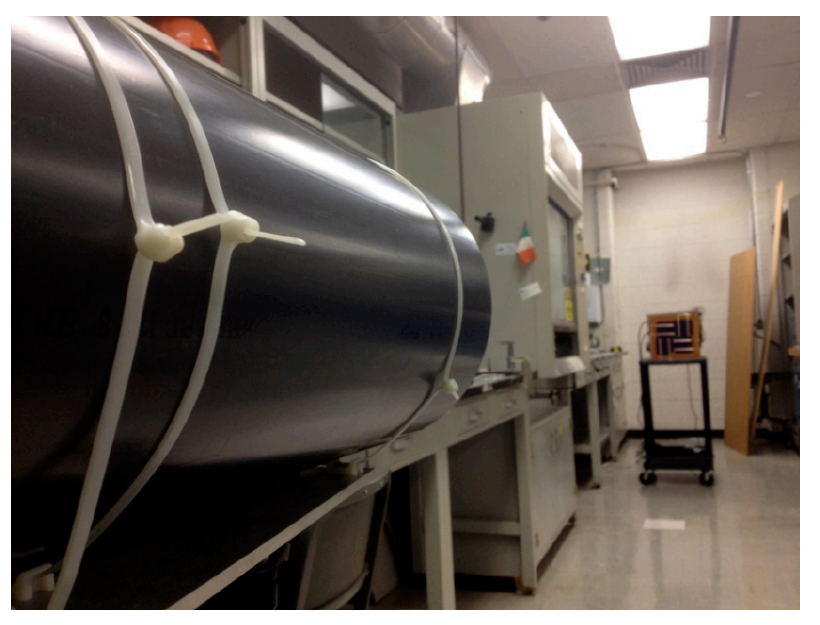}}
	\caption{Apparatus for optical power transmission. The LED and lens are inside the near tube, 
	                and the photovoltaic receiver is at the far end.}
\label{fig:fig_power_optical} 
\end{figure}

\begin{figure}[ht]
	\resizebox{\columnwidth}{!}{\includegraphics{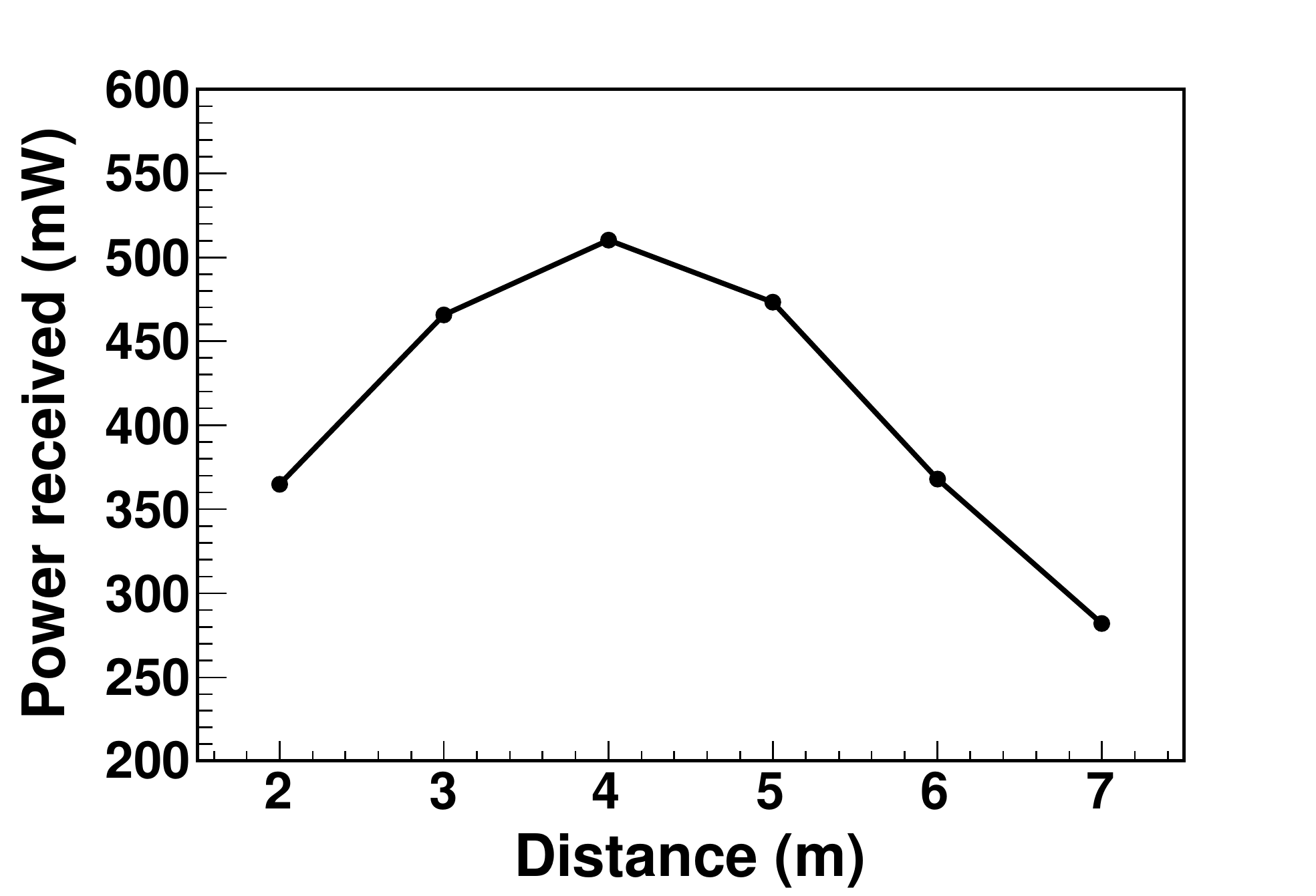}}
	\caption{Power received (mW) by the photovoltaic panel as a function of distance (m) from the optical source. 
	The optical source used was a $3.5$ W LED with a $940$ nm wavelength and a focusing lens.}
\label{fig:fig_optical_curve} 
\end{figure}

We tested both optical and RF power transfer methods. 
The optical demonstrator utilizes a high power light-emitting diode (LED)
that is collimated into an 8 inch diameter beam and is received by a photovoltaic (PV) panel, 
as shown in Fig.~\ref{fig:fig_power_optical}. 
The LED used is an OSRAM SFH 4751 with $3.5$ W optical output, operated at a maximum DC current of 1 A. 
The LED wavelength is $940$ nm which 
matches the peak efficiency of the Delsolar $156 \times 156 \; \mathrm{mm}^2$ photovoltaic cell used in our 
$312 \times 280 \; \mathrm{mm}^2$ PV panel array. Figure~\ref{fig:fig_optical_curve} shows the power received 
from this system as function of transmission distance. This test system met our prototype requirements of receiving 
$0.25$ W at 5 meters. 


\par
\begin{figure}[ht]
	\resizebox{\columnwidth}{!}{\includegraphics{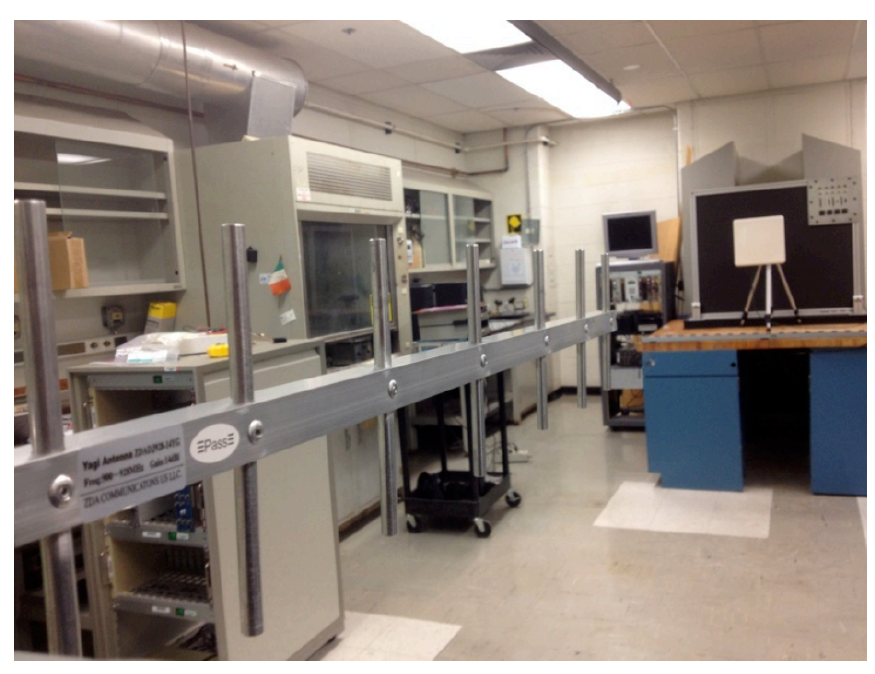}}
	\caption{Apparatus for RF power transmission.
	The RF transmitter ($14$ dBi gain Yagi antenna) is in the foreground,
	with the RF receiver ($11$ dBi gain patch antenna) at the far end.}
\label{fig:fig_power_RF}
\end{figure}

\begin{figure}[ht]
	\resizebox{\columnwidth}{!}{\includegraphics{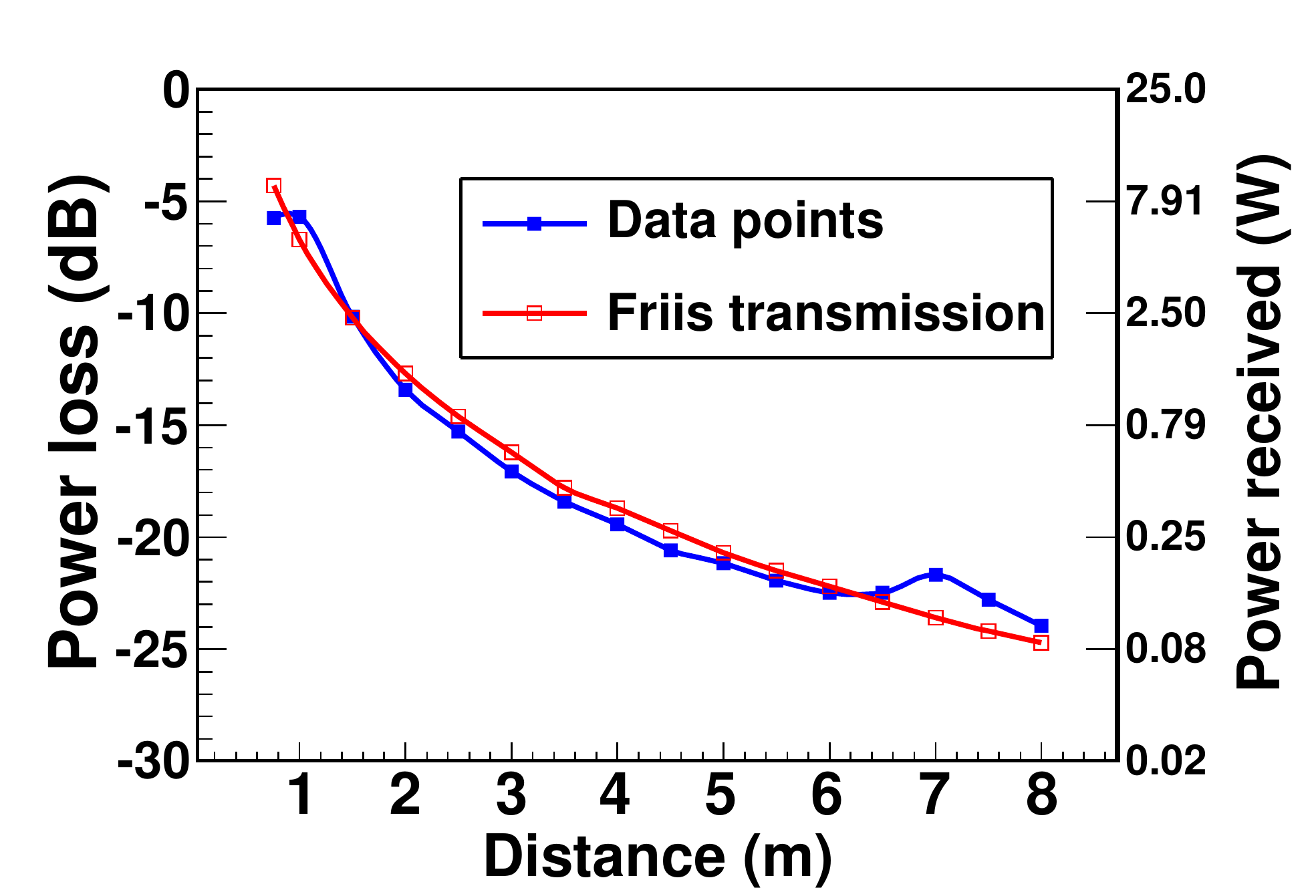}}
	\caption{Power loss (dB) in the RF power transmission test as a function of distance (m) from the source. 
	Solid dots are the measurements and open dots are the expectation from Friis transmission equation.
	The source is a $915$ MHz $10$ dBm transmitted 
	through a $14$ dBi gain Yagi antenna and received by an $11$ dBi gain patch antenna. 
	The scale on the right shows the power that would be received for a $25$ W ($44$ dBm) source.}
\label{fig:fig_RF_curve} 
\end{figure}

 The RF power transfer test consisted of a function generator driving a 14 dBi gain Yagi antenna at $915$ MHz 
with an output power of 10 dBm and being received by a 11 dBi gain patch antenna, as shown in Fig.~\ref{fig:fig_power_RF}. 
The measurements were taken in a large room at a significant height to minimize the scattering from surrounding objects.
The data from these tests, as a function of transmission distance are shown in Fig.~\ref{fig:fig_RF_curve}. 
The data points are in good agreement with the expectation from Friis transmission equation~\cite{friistrans},
which relates the ratio of power received $P_r$ to power transmitted $P_t$
for a given pair of antennas with gains $G_t$ and $G_r$, operating at an wavelength
$\lambda$, and separated by a distance $R$, as given in Eq.~\ref{eq:friiseq}. There is a rapid fall off of received power as the transmission distance increases. At 5 meters the power loss is $20$ dBm,
which requires a $25$ W source to receive $250$ mW, our target power.
\begin{equation}
	\frac{P_r}{P_t} = G_t G_r \left( \frac{\lambda}{4 \pi R} \right)^2 
\label{eq:friiseq}
\end{equation}

For this feasibility study, the optical system was the chosen prototype implementation. It provided a DC source
that is relatively easy to utilize. RF power transmission requires converting the RF power into a DC supply, 
which is commercially available but only at a $100$ mW receive power~\cite{ref_6}. The optical power transfer system that
we built met the power requirements for this demonstrator. 

RF power could be a better choice for a large detector since one source can power many front-ends.	
This greatly reduces the complexity and cost of the system. For example, at 5 meters the same power $\pm3$ dBm 
is being transmitted into a $2 \times 2 \; \mathrm{m}^2$ area. For a 50 W output, a single equivalent source could power 
as many as 40 front-ends depending on the packing density.


\section{Description of the Prototype System}
\label{sec_prototype}

\begin{figure}[htbp]
	\resizebox{\columnwidth}{!}{\includegraphics{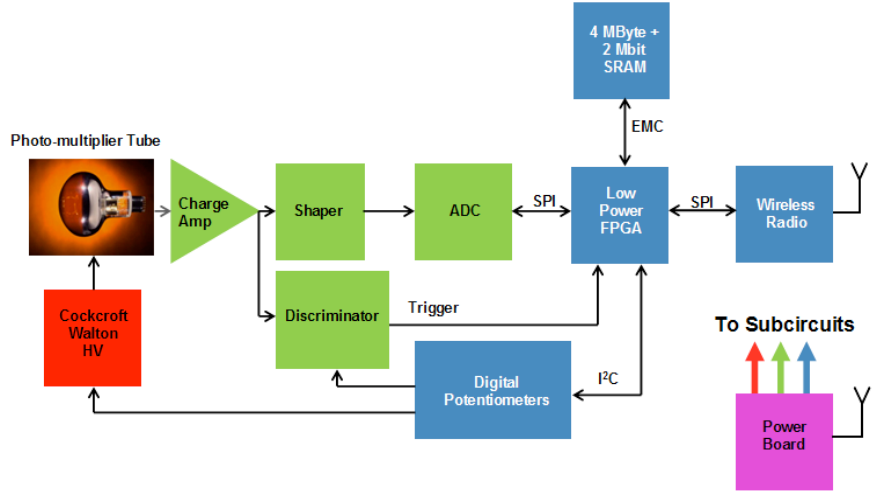}}
	\caption{Block diagram of the wireless prototype system.}
\label{fig:fig6} 	
\end{figure}

\begin{figure}[htbp]
	\resizebox{\columnwidth}{!}{\includegraphics{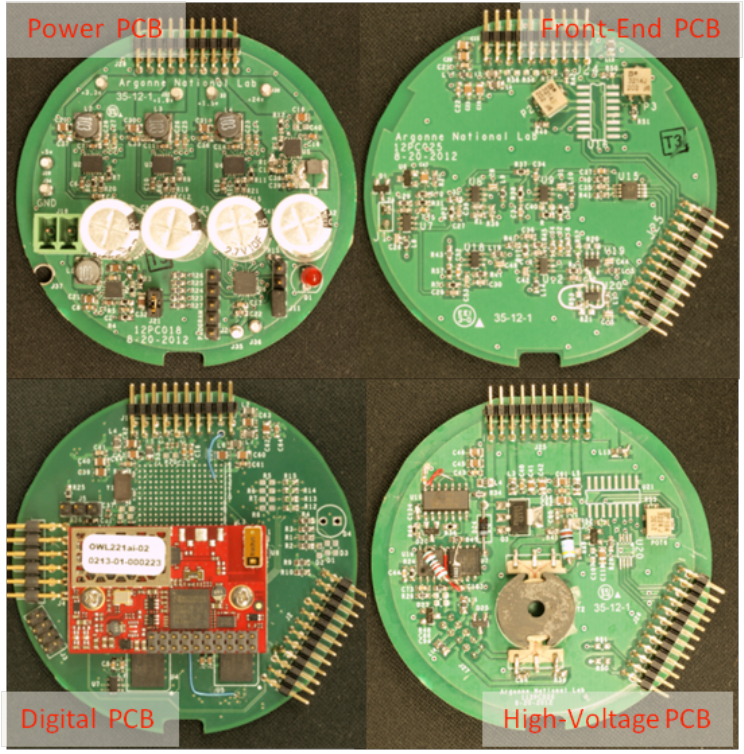}}
	\caption{Pictures of the four printed circuit boards in the wireless prototype system.
	Upper left figure shows the power board, upper right shows the front end, lower left shows the digital and lower right 
	shows the high voltage PCB. 
	}
\label{fig:fig7}
\end{figure}

\begin{figure}
	\resizebox{\columnwidth}{!}{\includegraphics{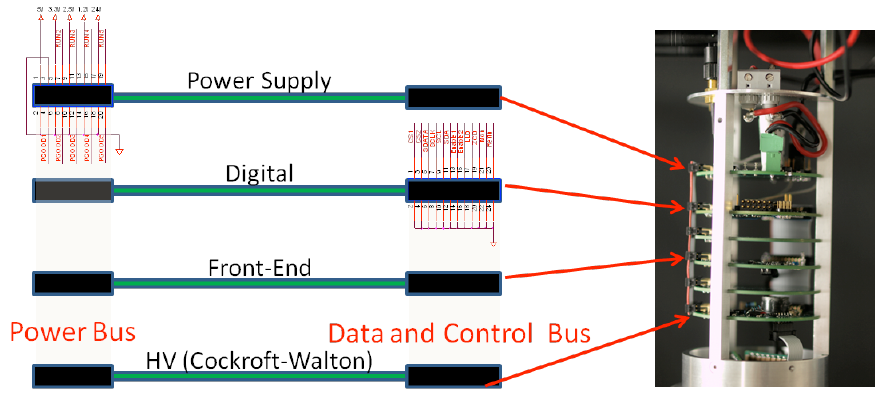}}
	\caption{Partial assembly of the system with the PMT.}
\label{fig:fig8}
\end{figure}

\begin{figure}
	\resizebox{\columnwidth}{!}{\includegraphics{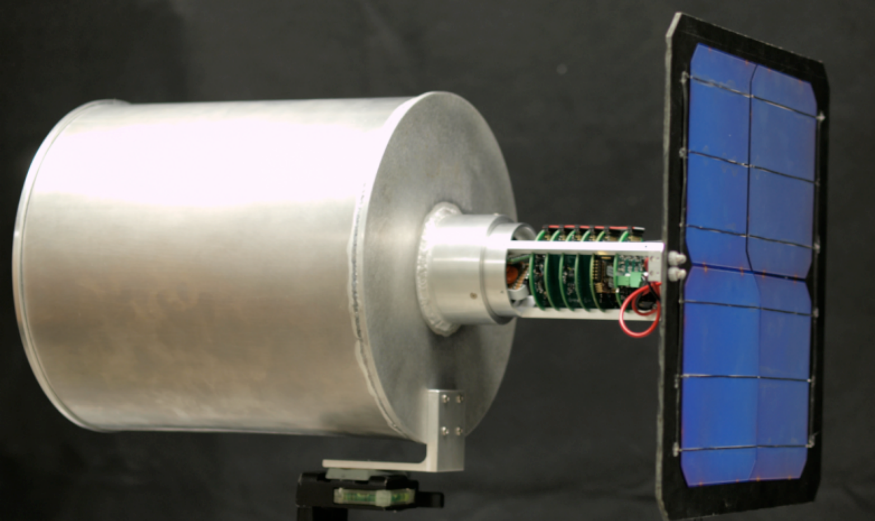}}
	\caption{Picture of the completed system. Shown with light tight PMT enclosure 
	and photovoltaic panel attached to a tripod (electronic enclosure open for reference).}
\label{fig:fig9}
\end{figure}

Figure~\ref{fig:fig6} shows a block diagram of the wireless demonstrator. The prototype system built is
comprised of $4$ boards that include: a board for generating high voltage for the PMT; a front-end board for shaping 
and digitization of the PMT signal; a digital board for processing the data and wireless data transmission; and a power 
board for receiving wireless power and generating the different voltages needed by the system. The system was 
designed to connect to a photovoltaic panel and send data wirelessly using 802.11n in the $5$ GHz band. 
Photos of the four boards in the PMT base are shown in Fig.~\ref{fig:fig7}. 
There is a power bus that connects the power PCB to the other three boards. 
A data bus connects the digital PCB to the front-end PCB and HV PCB. 
Physically they are arranged in the tube as shown in Fig.~\ref{fig:fig8}. 
The complete system with PMT is shown in Fig.~\ref{fig:fig9}. 



\subsection{Description of the Hardware}

\subsubsection{Power Board}

The power board uses a LTC 3105~\cite{ltc3105} switch mode step-up converter to turn the $1.85$ V received from 
the photovoltaic panel into one of the system voltages, $5$ V. The energy is stored in a series of ultra-capacitors~\cite{eshsr} 
with an effective total capacitance of $2.5$ F. The ultra-capacitors store energy locally and thereby reduce the peak power 
demand on the power system.
The $5$ V supply is used to generate $3.3$ V, $1.8$ V, and $1.5$ V with LTC 3104 step-down convertors~\cite{lt3104}, 
and $24$ V with a LT 3473 step-up converter~\cite{lt3473}. 
The digital board uses $3.3$ V, $1.8$ V, and $1.5$ V. The front-end board 
uses $3.3$ V. The high voltage board uses $5$ V and $24$ V. The overall conversion efficiency of the power conversion, 
relative to the photovoltaic input, is $70\%$.

\subsubsection{Digital Board}

The digital board houses the processor, wireless radio, and other digital ICs (external low-power SRAM, 
SPI flash, etc.) The wireless radio used is a commercial board from ConnectBlue, the cBOWL221a~\cite{connect_blue}. 
It is an 802.11n single stream module with an SPI interface. This module was chosen because of its low power 
consumption. With a low latency SPI bus running at $50$ MHz it supports a maximum payload transfer rate $35$ Mb/s. 
The radio offloads all of the $802.11$n protocol, but the TCP/IP or UDP/IP stack must be incorporated inside 
the processor. The digital processor used is a Microsemi A2F200 Smart Fusion~\cite{smart_fusion}. 
It combines a $100$ MHz hard ARM Cortex M3 processor, a relatively small FPGA, and includes 2 ADCs and 2 DACs. 
The physical hooks to use the ADCs and DACs for readout and control are included in the design but have not been 
exercised. The driver code for the radio is provided by ConnectBlue. The version that we used is written for use 
with embedded Linux. We chose to use FreeRTOS~\cite{rtos} because of the low overhead and relative ease in adaptation 
of the provided OS based driver. While this reduced the effort required for driver integration, it made hardware interrupt 
handling difficult to incorporate, so this was not implemented. As a result, the processing of each frame could begin 
only after the last data had been transferred to the wireless module. This reduced the overall throughput rate. 
A non OS-based single task process would more easily handle hardware interrupts, reduce SPI bus latency, 
and reduce processor overheads. For a large system, this would provide better performance, at the expense of 
a longer development time. The FPGA fabric is used for the discriminator logic, ADC readout, time-stamping, 
and event word generation. It directly accesses the external memory and stores the events using the same memory 
controller as used by the processor.

\subsubsection{High Voltage Board}

The high voltage board uses a standard Cockroft-Walton (CW) switching circuit to boost the $24$ V input voltage 
up to $2000$ V. The PMT is a 10-stage tube. Normally, the CW would have taps to connect directly to the dynodes. 
Unfortunately, the circuit was originally designed for a different PMT, and the change to the R7081HQE PMT made 
the connectivity incompatible. As a compromise, we implemented a resistive divider in the tube, but still retained 
the CW high voltage generation. While this worked acceptably, it caused the power consumption of the circuit 
to go up, as well as introducing rate limitations of the tube due to the relatively high values of resistance needed 
to make the circuit work. Nonetheless, we were able to run the circuit from wireless power, and the noise 
performance was sufficient to measure single photo-electrons.

\subsubsection{Front-end Board}
The front-end board includes a charge-sensitive preamplifier, shaping amplifier, programmable discriminators, 
and a $12$-bit ADC. The output of the charge sensitive amplifier connects to both a uni-polar and a bi- polar shaping 
amplifier. The uni-polar shaping amplifier connects to the AD7451 ADC~\cite{ad7451} and is used for digitizing 
the pulse-height. The bi-polar shaping amplifier is used as the input to pair of comparators, one of which is used 
for pulse-height discrimination and the other for timing. The comparator outputs connect to the A2F200 FPGA 
fabric (digital board) over the data and control bus. As discussed, the A2F200 FPGA fabric is used for 
the discriminator logic to control the ADC readout.

\subsection{Readout System and Process}

\begin{figure}[htbp]
	\resizebox{\columnwidth}{!}{\includegraphics{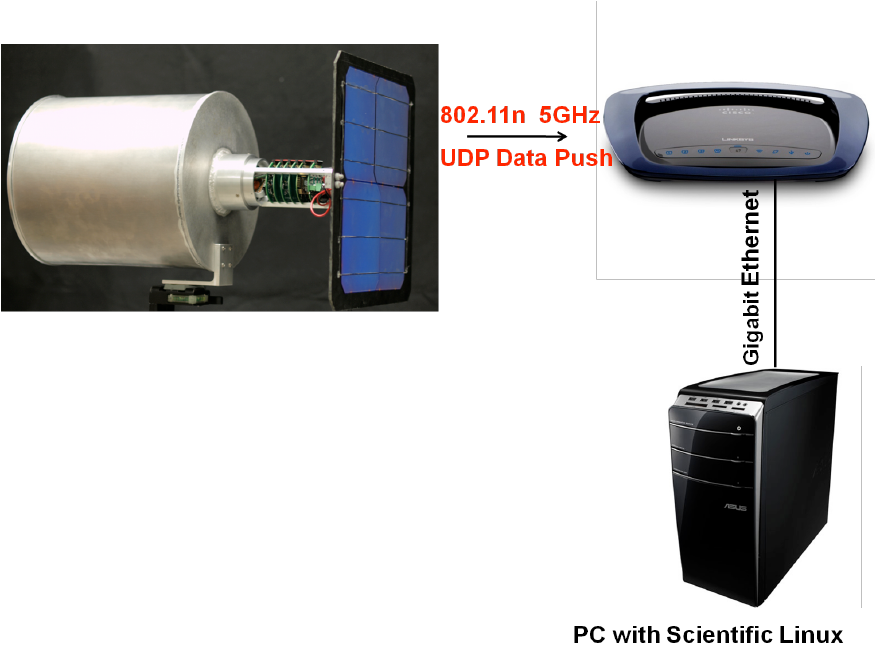}}
	\caption{Configuration of the prototype readout system.}
\label{fig:fig10}
\end{figure}

The prototype readout system is shown in Fig.~\ref{fig:fig10}. It consists of a Linux computer running Scientific 
Linux~\cite{slinux} and a commercial access point, a Cisco E3000~\cite{e3000} running DD-WRT firmware. 
The readout code was written to receive and store incoming UDP packets. For each asynchronous PMT trigger
the pulse height and time stamp are stored. Once per second the front-end transmits the data as a single UDP
packet to the server. A program running on the server receives and stores the UDP packets. Currently only data 
push from the front-end is implemented. As mentioned in section~\ref{sec_design}, implementing data pull is critical to extract 
the maximum throughput from multiple front-ends in a large detector.

\section{Performance}
\label{sec_perform}
The prototype system described above has been built and tested. It is powered wirelessly using the optical source received by the photovoltaic panel. All of the tests discussed here were performed with
the system running from wireless power. The performance achieved is summarized in Table~\ref{tab:table2}.
\begin{table*}
\centering
\caption{\label{tab:table2} Summary of initial goals and achieved performance for the wireless data acquisition system.}
\begin{tabular}{c|c|c}
     \hline\hline
Specification                                                  & Target                        & Performance   \cr   \hline
Total power consumption (@ $10$ kHz)     &   $250$ mW                   & $386$ mW           \cr  
\hspace{1cm} Digital                                    &   $120$ mW                   & $216$ mW          \cr
\hspace{1cm} Front-end                              &    $30$ mW                    &  $39$ mW          \cr
\hspace{1cm} HV                                          &    $80$ mW                    & $131$ mW          \cr   \hline
Maximum event rate                                    &    $10$ kHz                    &   $80$ kHz          \cr    \hline
Data transfer rate                                          &    $35$ Mb/s                   &   $11$ Mb/s          \cr    \hline
Bit Error Rate                                                 &  $<$ 1$\times$10$^{-12}$ & Dropped Packets  \cr    \hline\hline
\end{tabular}
\end{table*}

The power consumption of the system is $386$ mW, higher than the initial $250$ mW target. The primary causes of 
the increase are in the digital and the high voltage boards. In the case of the digital board, the target specification 
was based on spreadsheet power estimator, which does not accurately reflect actual operation. The high voltage 
board had to use a resistive divider as described earlier. Both of these problems are understood, and we feel 
that the original targets can be achieved in the next version of the design.
The system is capable of sending greater than $10$k events/s ($60$ kB/s). However, the critical parameter for 
performance in a large detector is the maximum burst transfer payload data rate. The specification of $35$ Mb/s 
is the maximum transfer rate of an $802.11$n single stream link, and this was our target. While this is achievable 
with the cBOWL221a radio, it does necessitate zero SPI bus latency and a $50$ MHz SPI clock. The maximum burst 
transfer rate that we achieved was $11$ Mb/s. The reason for this was twofold. First, the maximum SPI clock rate in 
the SmartFusion A2F200 is set by the master clock divided by $4$. This limits the maximum SPI clock rate to $25$ MHz. 
Second, latencies caused by FreeRTOS, the driver code, and the speed of the A2F200 processor, account for 
the remaining difference. The pin-compatible A2F500 supports a $50$ MHz SPI clock and will be implemented in 
the future. This simple change should yield $18$ Mb/s. Reducing the latency would require the elimination of 
FreeRTOS and driver optimization, as discussed earlier.
Our target bit error rate (BER) was less than $1\times10^{-12}$. A bit error rate test (BERT) program was written to analyze 
the performance. The prototype system has been run for many hours at a time with no bit errors observed in any
received packets. This is due in part to operating a relatively simple system, one broadcaster at a time, in 
the largely unused $5.5$ GHz band in our lab. It is also due to the forward error correction that is incorporated in 
the $802.11$n protocol. However, significant numbers of dropped packets were observed, on the order of $1$ in $2000$. It is
believed that this is due to the use of UDP. While more efficient than TCP, UDP does not have guaranteed transmission. 
We traced the problem to our access point, having observed the drop packets with the same frequency by using 
a computer to generate fake data. It may be possible to significantly reduce the systematic packet loss. The nature 
of UDP will require some consideration of error detection and recovery for a large system.

\begin{figure}[ht]
	\resizebox{\columnwidth}{!}{\includegraphics{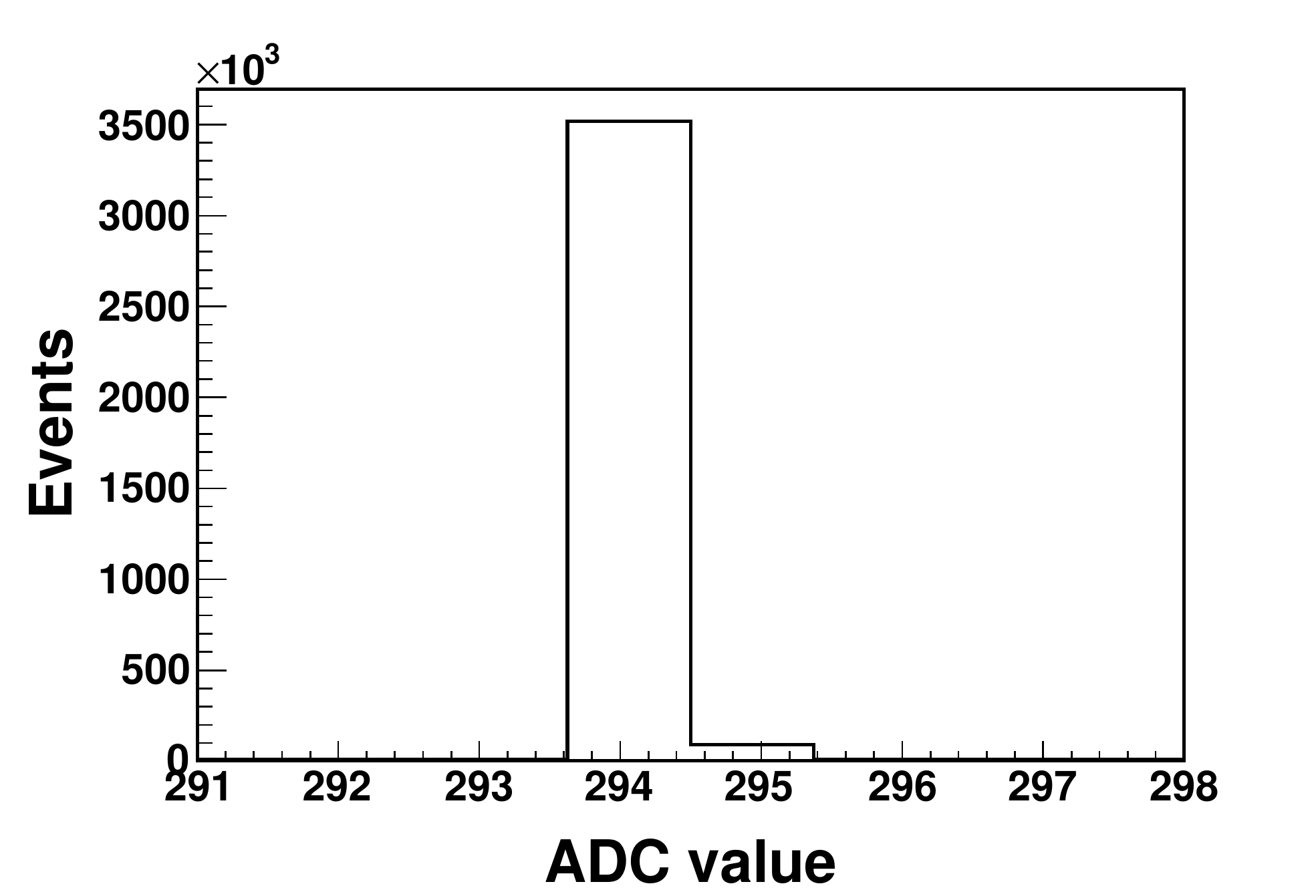}}
	\caption{\label{fig:fig_pedestal} Measurement of pedestal. The system is operated with conditions $10$k events/s, self-triggered, 
	HV on with no PMT, system powered wirelessly and wireless readout.}
\end{figure}

\begin{figure}[ht]
         \resizebox{\columnwidth}{!}{\includegraphics{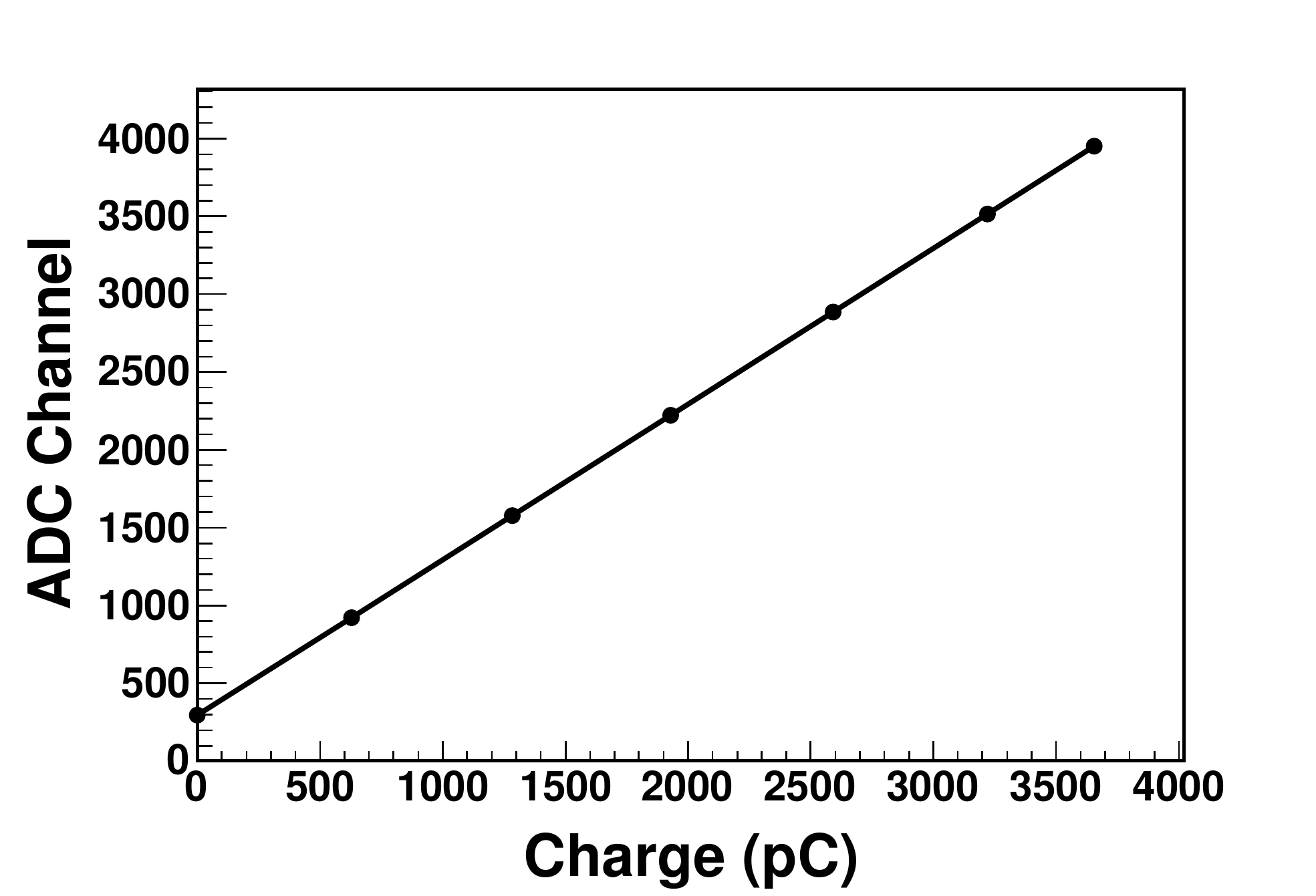}}
	\caption{\label{fig:fig_pulser} Response from electronic charge injection. The system is operated with conditions self-triggered, 
	PMT HV on, system powered wirelessly and wireless readout.}
\end{figure}

\begin{figure}[ht]
         \resizebox{\columnwidth}{!}{\includegraphics{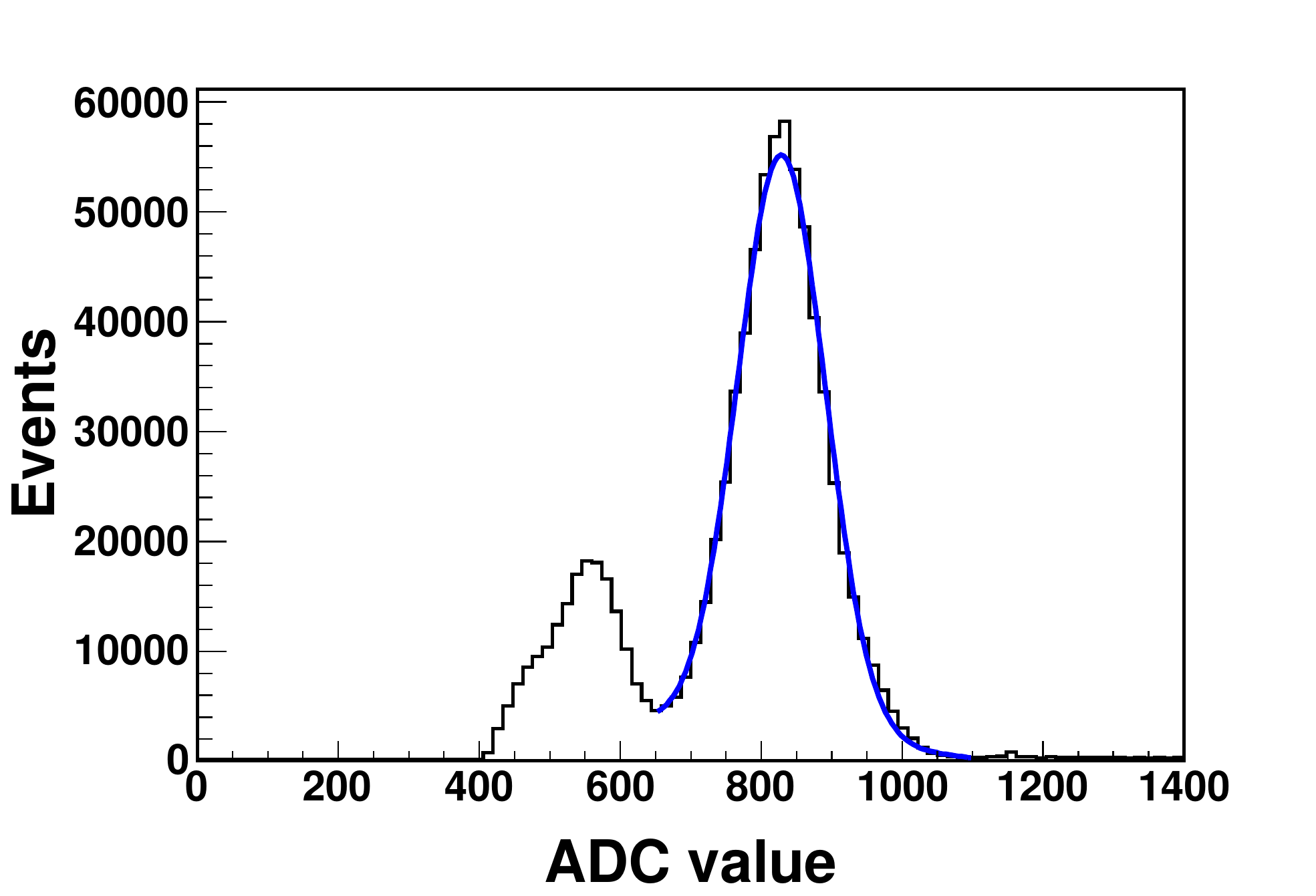}}
	\caption{\label{fig:fig_Am_source} Response from 60 keV $^{241}$Am source. The system is operated with conditions $1$k events/s, self-triggered, PMT HV $1300$V, system powered wirelessly and wireless readout. }
\end{figure}

\begin{figure}[ht]
         \resizebox{\columnwidth}{!}{\includegraphics{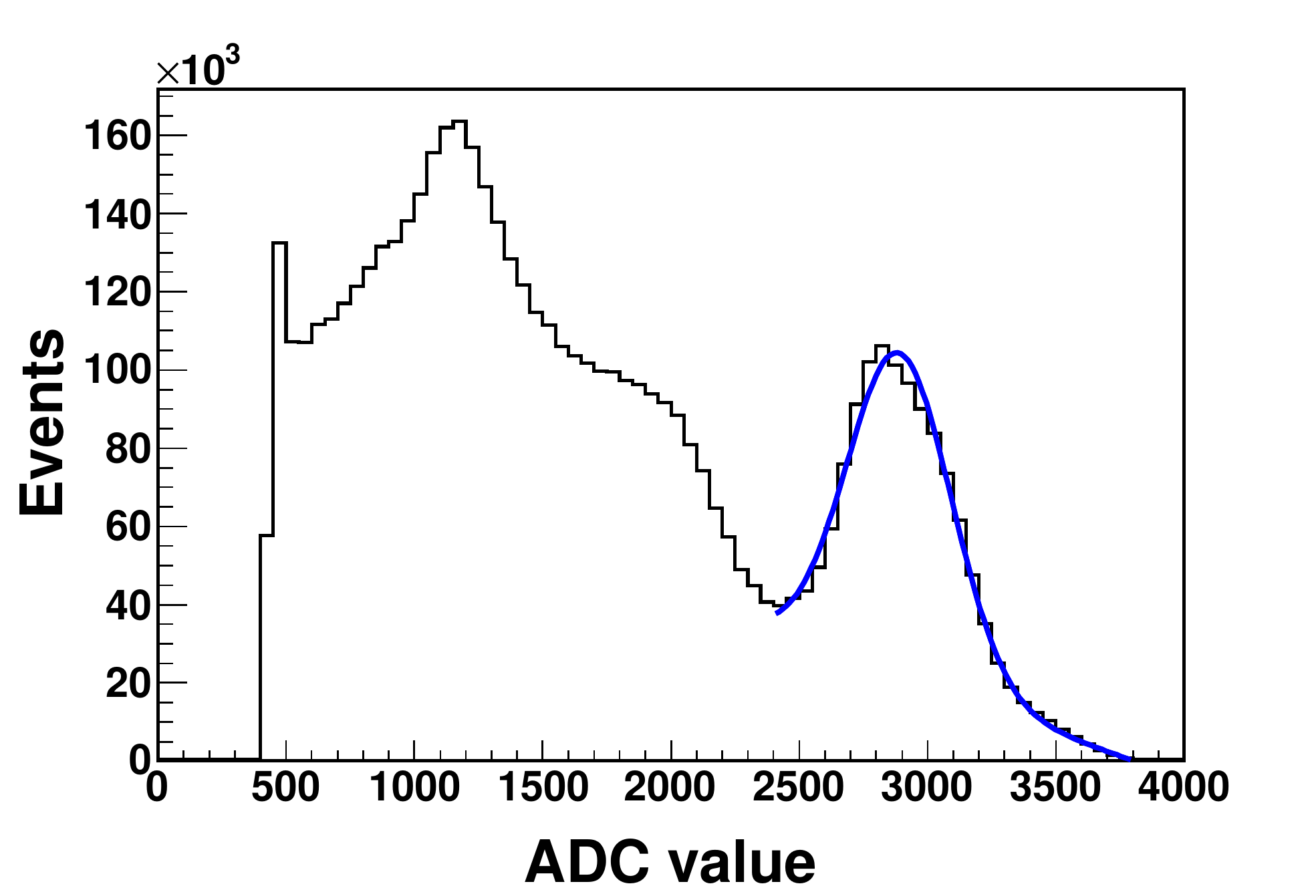}}
	\caption{\label{fig:fig_Cs_source} Response from 661.7 keV $^{137}$Cs source. The system is operated with conditions 250 events/s, self-triggered, 	 PMT HV $1200$V, system powered wirelessly and wireless readout.}
\end{figure}

We have collected data with the prototype system operated from wireless power and with wireless data readout. 
The front-end circuit exhibits very low noise, which is clear from the pedestal data in Fig.~\ref{fig:fig_pedestal}. 
The integral linearity over the full digitization range is quite good, better than $0.4\%$, as shown in Fig.~\ref{fig:fig_pulser}. 
To test the data acquisition capability, a sodium iodide crystal was attached to the PMT and tested 
with $^{241}$Am (Fig.~\ref{fig:fig_Am_source}) and $^{137}$Cs (Fig.~\ref{fig:fig_Cs_source}) sources. 
%
%

\section{Summary}

We successfully designed and built a wireless data acquisition system implemented in a photomultiplier 
tube base that operates from wireless power and sends data wirelessly. The power consumption, while 
slightly greater than our target, is still low enough to allow the system to operate from our optical power 
transfer system. The use of the pin-compatible A2F500 SmartFusion device should increase the burst 
transfer rate from $11$ Mb/s to $18$ Mb/s. Assuming $480$ Kb/s average data rate from our front-ends and assuming 
a conservative estimate of the latency between request for data and transmission of data of $10$ mS, the burst 
transfer rate achieved would support 18 and 27 front-ends for 11 and 18 Mb/s respectively. 
A system with $48$ access points would support 1296 front-ends.
Our future effort will address the SPI clock rate issue by using the A2F500, eliminating FreeRTOS, 
and reducing the SPI bus latency. This should reduce power consumption in the digital board. 
The high-voltage board will be redesigned to integrate correctly into the chosen PMT. 
In the longer-term, the intention is to use RF power transfer to facilitate the simplification of using 
one transmitter to power many receivers. Investigating the use of a custom ASIC for lower power 
operation of the front-end and Cockroft-Walton control circuitry would also be considered 
for a larger system.

\section{Acknowledgments}

{We acknowledge the support of Laboratory Directed Research \& Development from 
Argonne National Laboratory to carry out this project.
We would also like to thank Jeff Maus and Henric LindŽn of ConnectBlue for supporting our efforts to 
incorporate the cBOWL221a $802.11$n wireless radio working in our system. We especially thank 
Greg Mears of the Microsemi Corporation for his help with the SmatFusion system-on-chip device.








\end{document}